\documentstyle[11pt,fleqn,epsf]{article}

\newcommand{\beq}{\begin{equation}}
\newcommand{\eeq}{\end{equation}}
\newcommand{\beqa}{\begin{eqnarray}}
\newcommand{\eeqa}{\end{eqnarray}}
\newcommand{\sigr}{\sigma_{\rm r}}
\newcommand{\sigt}{\sigma_{\rm t}}
\newcommand{\tr}{t_{\rm r}}
\newcommand{\trhi}{t_{\rm rh,i}}
\newenvironment{reference}{\bigskip\bigskip\leftline{\Large\bf
References}\nopagebreak \begin{list}{}{\itemindent-\leftmargin \itemsep=0pt
\parsep=0pt}}{\end{list}}
\newcommand{\apj}{ApJ}
\newcommand{\apjs}{ApJS}
\newcommand{\mn}{MNRAS}
\newcommand{\pasj}{PASJ}

\newcommand{\ds}{\displaystyle}
\newcommand{\gtsim}{{\;{\ds\lower.7ex\hbox{$>$}\atop\ds\sim}\;}}
\newcommand{\ltsim}{{\;{\ds\lower.7ex\hbox{$<$}\atop\ds\sim}\;}}

\title{
\begin{flushright}
\normalsize
OU-TAP-30 \\
To appear in PASJ \\
August 1996
\end{flushright}
\ \\
Fokker-Planck Models of Star Clusters\\
with Anisotropic Velocity Distributions\\
II. Post-Collapse Evolution
}
\author{
Koji T{\normalsize AKAHASHI}
\thanks{Research Fellow of the Japan Society for the Promotion of Science}\\
{\it Department of Earth and Space Science,}\\
{\it Graduate School of Science, Osaka University,}\\
{\it Toyonaka, Osaka 560, Japan}\\
{\it E-mail: takahasi@vega.ess.sci.osaka-u.ac.jp}
}
\date{}

\begin{document}

\maketitle

%\begin{center}
%(Received 1996 January 5; accepted 1996 July 31)
%\end{center}

%\clearpage

\begin{abstract}

The evolution of spherical single-mass star clusters 
driven by two-body relaxation was followed 
beyond core collapse by numerically solving the orbit-averaged
Fokker-Planck equation in energy--angular momentum space.
The heating effect by three-body binaries was incorporated in the Fokker-Planck models.
The development of velocity anisotropy after the core collapse is discussed in detail.
The anisotropy in the outer regions continues to increase slowly after the collapse.
In clusters comprising a relatively small number of stars ($N \ltsim 10000$),
the post-collapse expansion is nearly self-similar 
and the anisotropy at each of inner Lagrangian radii is nearly constant.
In clusters comprising a larger number of stars ($N \gtsim 10000$),
gravothermal oscillations occur and the anisotropy at each of the inner radii oscillates with the core oscillations.
There is no qualitative difference in the nature of the gravothermal oscillations between the isotropic and anisotropic Fokker-Planck models.


\end{abstract}

\section{Introduction}\label{sec:intro}

\indent\indent

The dynamical evolution of globular clusters has been extensively investigated
[see Spitzer (1987) for a review].
In many of those investigations, Cohn's (1980) direct integration scheme for the Fokker-Planck (hereafter FP) equation was used as a main numerical tool.
Recently, many studies have been made particularly to reveal the {\it realistic} evolution of globular clusters and 
to compare the theoretical models with observations.
In such studies various effects were incorporated in numerical simulations: the stellar-mass spectrum, binaries, the galactic tidal field, stellar evolution, etc.
(e.g., Chernoff, Weinberg 1990; Drukier 1995).
On the other hand,
the anisotropy of the velocity distribution was almost always neglected,
although it is obvious that the anisotropy develops at least in the halo
(it is expected that the radial velocity dispersion exceeds the tangential one).
This neglect was mainly due to a numerical difficulty involving anisotropic FP models.
The evolution of anisotropic clusters can be described by a two-dimensional (2D) orbit-averaged FP equation in energy--angular momentum space (Cohn 1979),
while the evolution of isotropic ones can be described by a one-dimensional (1D) orbit-averaged FP equation in energy space (Cohn 1980).
The direct integration code for the 2D FP equation had a numerical problem 
in that the energy conservation was insufficient to continue the run 
beyond a factor of $10^3$ increase in the central density (Cohn 1979; Cohn 1985).
On the other hand, an integration of the 1D FP equation can be performed with much higher numerical accuracy (Cohn 1980), 
largely due to the adoption of the Chang-Cooper finite differencing scheme (Chang, Cooper 1970).

Recently, Takahashi (1995, hereafter Paper I) has developed a numerical method for solving the 2D FP equation.
The method is essentially the same as Cohn's (1979) method.
A main difference between the two methods exists concerning discretization schemes of the FP equation.
Cohn (1979) used a finite-difference scheme in which
simple centered-differencing was adopted for the spatial discretization.
Cohn (1985) reported that he investigated several heuristic
generalizations of the Chang-Cooper scheme, and that all of these improved
the energy conservation,
though the details of these schemes were not explained.
In Paper I, two different discretization schemes were employed:
one was a finite-difference scheme where the Chang-Cooper scheme is simply applied for only the energy direction;
the other was the finite-element scheme, where the test and weight functions
implied by the generalized variational principle (Inagaki, Lynden-Bell 1990; Takahashi 1993) are used.
Using those schemes, the gravothermal core collapse was followed until the central density increased by a factor of $10^{14}$ with a 1\% numerical accuracy 
concerning the total-energy conservation.
This was a big advance compared with previous calculations;
the central density growth factors in the calculations of Cohn (1979) and Cohn (1985) were $10^3$ and $10^6$, respectively.
It should be noted that 
a numerical error originates not only in the integration of the FP equation, 
but also in other calculation procedures, e.g., the calculation of the diffusion coefficients and the potential-recalculation steps.
It should also be noted that 
2D FP calculations require a rather large computational time (see section \ref{sec:calc}).
Thus, ten years ago it was not easy to perform such calculations 
as those which we present here.
Besides the FP models, 
anisotropic gaseous models of star clusters have recently been successfully 
applied (e.g., Giersz, Spurzem 1994; Spurzem 1996).

In Paper I, the pre-collapse evolution of single-mass clusters was studied.
In particular, Paper I revealed the evolution during self-similar phases of core collapse in anisotropic clusters.
The density profile left outside the collapsing core is the same as that
in isotropic clusters; i.e. $\rho \propto r^{-2.23}$.
In the self-similar regions, a slight velocity anisotropy exists:
i.e. $\sigt^2/\sigr^2 = 0.92$, where $\sigr$ and $\sigt$ are the one-dimensional radial and tangential velocity dispersions, respectively.
The core collapse rate, $\xi \equiv \tr(0)d\ln\rho(0)/dt$, where $\tr(0)$ and $\rho(0)$ are the central relaxation time and density, is $\xi = 2.9\times10^{-3}$, which is 19\% smaller than the value of $\xi = 3.6\times10^{-3}$ for an isotropic model.
That is, the core collapse proceeds slightly more slowly in the anisotropic model than in the isotropic model.
When the initial model is Plummer's model,
the collapse occurs at time $17.6\,t_{\rm rh,i}$ in the anisotropic model and
$15.6\,t_{\rm rh,i}$ in the isotropic model, where $t_{\rm rh,i}$ is the initial half-mass relaxation time.
The halo soon becomes dominated by radial orbits, even if the velocity distribution is initially isotropic everywhere.
The ratio of the radial velocity dispersion to the tangential one 
increases monotonically as the radius increases.

Following Paper I, this paper examines the post-collapse evolution of single-mass clusters.
The effect of three-body binaries is incorporated into FP models as a heat source.
We are particularly interested in the development of the anisotropy in the halo after the core collapse.
Does the anisotropy continue to increase even after the collapse, 
or come to be constant ?
We are also interested in whether there are any differences concerning the nature of gravothermal oscillations between isotropic and anisotropic models.
In section 2, the models and numerical methods are described.
In section 3, the calculation details are described and the numerical accuracy is discussed.
Section 4 presents the results of the calculations.
The conclusions and discussion are given in section 5.

\section{The Models and Methods}\label{sec:model}

\subsection{\normalsize \it Fokker-Planck Models}

%\indent\indent
\indent

We consider the collisional evolution of spherical single-mass star clusters in dynamical equilibrium.
In such clusters the distribution function of stars, $f$, is a function of the energy per unit mass $E$, the modulus of the angular momentum per unit mass $J$, and time $t$; i.e. $f=f(E,J,t)$.
The evolution of $f$ due to the two-body relaxation can be described by an orbit-averaged FP equation in $(E,J)$-space (Cohn 1979).
Numerical integration of the FP equation is performed in the same manner as in Paper I,
but a binary heating term is included in the equation.

For problems concerning post-collapse evolution, 
we must specify the number of stars in the cluster, $N$, and the numerical constant, $\mu$, in the Coulomb logarithm $\ln (\mu N)$ (e.g. Spitzer 1987, p30).
In all of the calculations we adopted $\mu=0.11$, 
which was obtained by Giersz and Heggie (1994a) for the pre-collapse evolution of single-mass clusters.
We note that Giersz and Heggie (1994b) found a smaller value of $\mu$ 
for the post-collapse evolution (their best value was $\mu=0.035$).
However, we fixed the value of $\mu$ throughout all evolutionary phases.
A small difference in $\mu$ does not seriously affect the nature of cluster evolution.
Although the determination of an appropriate value of $\mu$ is an interesting subject in collisional stellar dynamics,
it was beyond the scope of this study.
A future careful comparison between $N$-body, gaseous, and FP models may give further information concerning the Coulomb logarithm.

\subsection{\normalsize \it Three-Body Binary Heating}

%\indent\indent
\indent

The three-body binary heating rate per unit mass is given by
\beq
\dot{E}_{\rm b}=C_{\rm b} G^5 m^3 \rho^2 \sigma^{-7}  \label{eq:lhr}
\eeq
(Hut 1985), where $m$ is the stellar mass, $\rho$ the mass density, $\sigma$ the one-dimensional velocity dispersion, and $C_{\rm b}$ a numerical coefficient.
In this paper we choose the standard value of $C_{\rm b}=90$.
The local heating rate (\ref{eq:lhr}) is orbit-averaged as
\beq
\langle \dot{E}_{\rm b} \rangle_{\rm orb} 
= \left. \int_{r_{\rm p}}^{r_{\rm a}} \frac{dr}{v_{\rm r}} \dot{E}_{\rm b}
\right/ \int_{r_{\rm p}}^{r_{\rm a}} \frac{dr}{v_{\rm r}} \,, \label{eq:oahr}
\eeq
where $v_{\rm r}=\left\{2[\phi(r)-E]-J^2/r^2\right\}^{1/2}$ is the radial velocity of a star of energy $E$ and angular momentum $J$ at radius $r$,
and $r_{\rm p}$ and $r_{\rm a}$ are the pericenter and apocenter radii of the star, respectively.
The orbit-averaged heating rate (\ref{eq:oahr}) is added to the usual 
first-order diffusion coefficient 
$\langle \Delta E \rangle_{\rm orb}$ (cf. Cohn et al. 1989).
Furthermore, we assume that the scattering by binaries does not produce the net changes of the scaled angular momentum $R$, 
i.e. $\langle\dot{R}_{\rm b}\rangle_{\rm orb}=0$.
Here, $R$ is defined as
$R=J^2/J_{\rm c}^2(E)$, where $J_{\rm c}(E)$ is the angular momentum of a circular orbit of energy $E$.

\section{Numerical Calculations}\label{sec:calc}

\indent\indent

Plummer's model (e.g. Spitzer 1987, p13) was chosen as the initial cluster model,
where the velocity distribution is isotropic everywhere.
Test calculations were carried out using both the finite-difference and
finite-element codes described in Paper I\@.
 In calculations of the pre-collapse evolution, 
the two codes achieved similar numerical accuracy concerning the total energy and mass conservation,
and the results obtained using them were generally in good agreement (Paper I).
In the present calculations of the post-collapse evolution, 
the results obtained by the two codes were also generally in good agreement.
However, the numerical error in the energy conservation was considerably larger in the case of the finite-element code.

We note that the total energy of the cluster cannot be conserved in these calculations, but should increase,
because the binary heating is taken into account.
The total energy should increase by the amount of energy input.
To check the numerical error, the energy input was recorded during the calculations.
The energy input at each time step may be calculated by integrating
the product of the binary heating rate [equation (\ref{eq:oahr})] 
and the distribution function over energy--angular momentum space.
The cumulative energy input is summed up over the time steps of the run.
There must be some degree of inaccuracy inherent in this estimation for the energy input.
However, this estimated energy input and the actual energy increase resulting from the FP-integration should be in agreement within some degree of numerical accuracy;
the degree of the agreement becomes better as the mesh sizes and the time step size become smaller.
When we estimated the numerical error in the energy conservation,
we assumed that the energy input estimated as above was exact.

For example, 
at the end of the calculation for $N=5000$ with the finite-difference code (see figure 1a), 
the relative energy error, which is defined as the ratio of the amount of the energy change due to numerical error to the initial total energy, was about 2\%;
at the end of a corresponding calculation with the finite-element code, the relative energy error was about 12\%.
There was a systematic energy drift during calculations of the post-collapse evolution in both the finite-difference and finite-element codes.
This error arose mainly from the FP-calculation steps.
In fact, in one FP step, the actual energy increase was always slightly smaller than the expected increase due to binary heating.
The degree of this disagreement was much larger in the finite-element code than in the finite-difference code.
An energy error arose also from the Poisson-calculation steps.
However, the sign of the energy change in one Poisson step was nearly random,
and the sum of the changes was small. 
Therefore, the energy error stemming from the Poisson steps does not 
contribute very much to a cumulative error.

The reason why the accuracy of the finite-element code for the energy conservation is not very good for the post-collapse calculations is not yet clear.
One way to improve the accuracy is to increase the mesh numbers, especially for the energy.
In fact, the energy error decreased as the energy-mesh number increased,
although the accuracy of the finite-difference code was better with the same mesh number.
Another promising way to improve the accuracy of the finite-element scheme is to use higher-order basis functions (see appendix 2 of Paper I).
In the present code two-dimensional piecewise bilinear polynomials are used as the basis functions.
The use of higher-order basis functions, however, introduces rather complicated computational procedures, and, as a result, a larger computational time.
In addition, we can obtain reasonably good accuracy with the finite-difference code.
Thus, we did not try higher-order basis functions in the present work.

As a result, we adopted the finite-difference code for the calculations presented in this paper because of its higher numerical accuracy in energy conservation.
The results of calculations for $N=$ 5000, 10000, 20000, and 40000 are shown in section \ref{sec:result}.
We denote the number of grid points in $X$, $Y$, and $r$ by $N_X$, $N_Y$,
and $N_r$, respectively.
[Variables $X$ and $Y$ are used instead of $E$ and $R$ in the code (Paper I).]
In these calculations,  we set $N_X=151$, $N_Y=35$, and $N_r=91$.
The radial grid was constructed between $10^{-7}~r_0$ and $10^2~r_0$,
where $r_0$ is a length scale parameter of Plummer's model.
We carried out several test calculations with other sets of grid numbers,
and confirmed that the results converged.
The relative energy errors at the ends of the calculations shown in figure 1 were
2.3\% ($N=5000$), 2.5\% ($N=10000$), 1.1\% ($N=20000$), and 0.1\% ($N=40000$).
In the calculation of $N=40000$, the energy error reached its maximum (0.5\%) at the first collapse time.
However, since the sign of the energy error changed with the core oscillations,
there was some cancellation in the cumulative error.
The relative mass errors were
0.85\% ($N=5000$), 0.87\% ($N=10000$), 0.51\% ($N=20000$), and 0.57\% ($N=40000$).

One inevitable disadvantage of the 2D FP model relative to the 1D FP model is that 2D calculations take a much larger computational time than do 1D calculations.
One may expect that 2D calculations require about $N_Y$ times as large a
computational time as do 1D calculations.
In fact, however, it can happen that the computational time of 2D calculations increases faster than as $N_Y$.
The computational time required to solve a linear matrix equation for a discretized FP equation is not negligible, but, rather, a few tens of a percent of the
total computational time in 2D calculations.
In 1D calculations, in contrast, it is almost negligible compared with the
total computational time, because the matrix is tridiagonal and can be inverted very easily.
In 2D calculations, the matrix is a band matrix whose half-bandwidth is about $\min (N_X,N_Y)$, or $N_Y$ in our cases. 
We can choose various direct or iterative schemes for solving the matrix equation.
In some direct schemes for band matrices, the number of required operations varies as $N_X N_Y^3$ for large $N_X$ and $N_Y$.
A kind of conjugate gradient method (iterative method) was actually used in our 2D FP code.
The number of operations varies as $N_X N_Y$ for this method.
We found by experience that the computational time required by a 2D calculation with our code is about $2N_Y$-times larger than that required by a corresponding 1D calculation (with the same $N_X$).
Most of the numerical calculations were performed on a HP 9000/715 workstation
(at 50~MHz clock cycle).
For example,
2D FP calculations for $N=$ 5000 and 40000 (cf. figure 1)
required about 29 and 140 hours of CPU time on this machine, respectively.
The total numbers of potential-recalculation time steps 
(the FP time-step size was 1/10 of the potential-recalculation time-step size) 
in these runs were 3000 and 15000, respectively, 
and thus the 2D FP code required about 34 CPU sec per step.

\section{Results}\label{sec:result}

\indent\indent

The results are presented in standard units such that
$G=M=1$ and ${\cal E}_{\rm i}=1/4$, where $G$ is the gravitational
constant, $M$ is the total mass,
and ${\cal E}_{\rm i}$ is the initial total binding energy.
The time is usually measured in units of the initial half-mass relaxation time $t_{\rm rh,i}$ (Spitzer 1987, p40).

%\begin{center}
%------------------------------\\
%Fig. 1a,b,c,d\\
%   Central density evolution
%------------------------------\\
%\end{center}

Figure 1 shows the evolution of the central density $\rho(0)$ for the cases of $N=$ 5000, 10000, 20000, and 40000 for the 1D and 2D models.
For each $N$, the features of the evolution in the 1D and 2D models are very similar.
An apparent difference between the two models exists in the core collapse times.
For every $N$, the core collapse (or bounce) occurs slightly earlier in the 1D model than in the 2D model, as found in Paper I\@.
 Although an intuitive explanation for a slower collapse in the 2D (i.e. anisotropic) models is given in Paper I (see also Louis 1990),
a more convincing proof for it is desirable.
There is 
a possibility that the slower collapse may be due to the numerical inaccuracy.
We may be able to test this possibility simply by repeating the calculations with finer grids.
We found that the collapse time was not affected by increasing the grid numbers.
This fact supports the conclusion that the slower collapse rate in the anisotropic models is real.
We also note that it is uncertain whether adopting the isotropic distribution function to calculate the diffusion coefficients (Paper I) has any noticeable effects on the collapse rate.

The core expansion after the core collapse is stable for $N=5000$,
and overstable for $N=10000$.
(For $N=10000$, the core expansion in the 1D model is really overstable, though the growth of the instability is slower than in the 2D model.)
For $N=20000$, the core expansion is unstable;
the central density oscillates chaotically with a large amplitude; that is, 
gravothermal oscillations (Bettwieser, Sugimoto 1984) occur.
The gravothermal oscillations also occur for $N=40000$ with a larger amplitude than for $N=20000$.
Such a change in the nature of the post-collapse core evolution from monotonic expansion to chaotic oscillations with increasing $N$ was discussed in detail by Goodman (1987), Heggie and Ramamani (1989), Breeden et al. (1994), as well as Breeden and Cohn (1995), where isotropic models were used.
Spurzem (1994) presented long-lasting gravothermal oscillations
in his anisotropic gaseous model.
We see no qualitative difference concerning the features of the gravothermal oscillations between the 1D and 2D models.
The amplitudes and periods of the oscillations, 
and the appearance of multiple-peaks in the two models are similar.
This is a reasonable result,
because the stability of the core expansion is determined by the degree of central concentration (Goodman 1987).
Furthermore, the velocity distribution is isotropic in the core, even in anisotropic models.
It is interesting that Spurzem (1994) suggested that gravothermal oscillations are more regular in the anisotropic model than in the isotropic one.

%\begin{center}
%------------------------------\\
%Fig. 2a,b\\
%   Lag. radii, N=5000
%------------------------------\\
%\end{center}

Figure 2a shows the evolution of the Lagrangian radii containing
1, 2, 5, 10, 20, 30, 40, 50, 75, and 90\% of the total mass for $N=$5000.
The core radius $r_{\rm c}$ is also plotted;
it is defined as
\beq
r_{\rm c} \equiv \left[\frac{3v_{\rm m}(0)^2}{4\pi G\rho(0)}\right]^{1/2}
\eeq
(Spitzer 1987, p16),
where $v_{\rm m}(0)$ is the total velocity dispersion at the center.
Because of the difference in the collapse time, 
a comparison between the 1D and 2D models concerning the evolution of the spatial structure is somewhat complicated.
Thus, we plot figure 2b, where the time of the 1D calculation is scaled so that the collapse time in the 1D model should coincide with that in the 2D model.
Concerning the evolution before the core bounce, we see small difference
between the two models in figure 2b,
except for the 90\% radius.
After the core bounce, the core-halo structure is more developed in the 2D model;
that is, the 2D model has more concentrated inner Lagrangian radii and more extended outer radii.
We note that there is no big difference between the two models in the evolution of the half-mass radius.
The evolution of the half-mass radius is, roughly speaking, determined by the change in the total energy when the total mass is conserved.
In fact, the histories of the total energy changes in the two models are similar, if the time is scaled as in figure 2b.
In this respect, the coincidence of the evolution of the half-mass radius is reasonable.
The effects of the development of the anisotropy on the density is apparent in the outer half-mass region.
This is a consequence of the development of radial orbits
that the outer Lagrangian radii are more extended in the 2D model.
The more concentrated inner Lagrangian radii are a necessary reaction to that.
However, the evolution of the core radius in the two models is 
again almost identical (if the time of the 1D models is scaled).

Figure 2b shows that
the post-collapse evolution well after the core bounce seems to be self-similar in both the 1D and 2D models;
all of the Lagrangian radii as well as the core radius expand nearly self-similarly.
A simple argument gives a self-similar expansion law, $r \propto t^{2/3}$, for isolated clusters with no mass-loss (H\'enon 1965; Goodman 1984).
Our 2D model as well as our 1D model is consistent with this law.
In the cases of other $N$'s, the evolution of the outer Lagrangian radii is similar to that in the case of $N=5000$.
When gravothermal oscillations occur, the inner Lagrangian radii oscillate, 
while the mean trend of these radii is also an expansion (cf. figure 5).

%\begin{center}
%------------------------------\\
%Fig. 3a,b \\
%   Anisotropy: N=5000
%------------------------------\\
%\end{center}

Figures 3a and 3b show the evolution of the anisotropy parameter $A$, 
\beq
A \equiv 2-2\frac{\sigt^2}{\sigr^2} 
\eeq
at the 1, 2,..., and 90\% Lagrangian radii, for the case of $N=5000$.
During the core collapse, $A$ increases at every Lagrangian radius. 
Even in the very inner regions (e.g. at the 1 and 2\% radii) the anisotropy increases at advanced stages of the collapse.
Just after the core bounce, the anisotropy at each of the inner (1--20\%) Lagrangian radii decreases rapidly.
This is due to a rapid core expansion.
The core expansion is faster than the expansion of the Lagrangian radii located outside the core at that time.
Then, the radial velocity dispersion decreases faster than the tangential one outside of the core,
because the former is influenced more by the core condition.
This is an exactly opposite process to that occurring during the core collapse.
After the rapid core expansion phase,
the cluster expands nearly self-similarly (as mentioned above),
and the anisotropy at each inner Lagrangian radius settles to roughly a
constant value.

The anisotropy at the outer Lagrangian radii continues to slowly increase after the core bounce.
In figure 3b we can see that the curve of the anisotropy at the 90\% radius flattens at late times.
This is partly because $A$ cannot exceed two, by definition.
In any case, it is true that the rate of the anisotropy increase at the outer radii slows down.
The development of the anisotropy in the outer regions is a consequence of the emergence of radial-orbit stars which have gained energy as the result of relaxation in the inner regions (Paper I).
Therefore, we expect that the rate of the anisotropy increase is related to the relaxation time in the inner regions.

%\begin{center}
%------------------------------\\
%Fig. 4a,b \\
%   Anisotropy: N=5000, tau
%------------------------------\\
%\end{center}

Figure 4 shows the evolution of the anisotropy at the Lagrangian radii for
$N=5000$ as a function of the elapsed number of actual central relaxation times,
\beq
\tau (t) \equiv \int_0^t \frac{dt'}{t_{\rm rc}(t')}  \label{eq:tau}
\eeq
(Cohn et al. 1989), where $t_{\rm rc}$ is the central relaxation time.
The core bounce occurs at $\tau \approx 2400$.
This figure indicates that the anisotropy at the outer Lagrangian radii increases roughly linearly with $\tau$ after a rapid increase at the very initial stages ($\tau \ltsim 1000$).
While we see in figure 3b that the rate of increase of the anisotropy at the outer radii change sharply at the time of the core bounce,
we do not see any such sharp changes in figure 4b.
These facts tell us that the slowing down of the increase rate of the anisotropy after the core bounce appearing in figure 3a is mainly due to 
the fact that the central relaxation time becomes longer and longer as the cluster expands.

In the statistical data from $N$-body simulations for $N=1000$ by Giersz and Spurzem (1994, figure 11), 
we can see that the anisotropy at outer Lagrangian radii reaches its maximum around the core bounce time, and then decreases.
Such a decrease does not occur in our 2D FP models.
Giersz and Spurzem (1994) as well as Giersz and Heggie (1994b) argued that the anisotropy in the outer regions is determined (at least partially) by binary activity:
interactions of binaries with single stars, and the expulsion of stars and binaries from the core to the outer parts of the system.
Such effects are not completely included in our models, but binaries only play the role of a continuous heat source.
The effects of binaries on the anisotropy may be important for small-$N$ systems
and responsible for the fact that the anisotropy reaches its maximum around the core bounce in the 1000-body model. 
For $N=10000$ clusters, there is a good agreement in the evolution of the anisotropy in the outer regions 
between the 2D FP and $N$-body models (Spurzem 1996; Takahashi 1996).
It is not clear whether or not the anisotropy at the outer radii decreases after the core bounce in a 10000-body simulation (by Spurzem), 
because the simulation has not yet been continued enough beyond the core bounce.

%\begin{center}
%------------------------------\\
%Fig. 5\\
%   Lag: N=20000
%------------------------------\\
%\end{center}

%\begin{center}
%------------------------------\\
%Fig. 6a,b \\
%   Anisotropy: N=20000
%------------------------------\\
%\end{center}

Figure 5 shows the evolution of the Lagrangian radii and the core radius for 
$N=20000$.
In this case, the post-collapse evolution of the inner Lagrangian radii is not self-similar, but gravothermal oscillations occur.
However, the outer Lagrangian radii expand nearly self-similarly after the first collapse, just as in the case of $N=5000$.
The evolution of the anisotropy $A$ for $N=20000$ is shown in figure 6.
The anisotropy at each of the inner Lagrangian radii reaches a higher value at the first collapse time for $N=20000$ than for $N=5000$.
This is because the core collapse proceeds to more advanced stages
and the anisotropy penetrates into more inner regions (cf. figure 3 of Paper I) for $N=20000$.
After the first core collapse the anisotropy at the inner Lagrangian radii oscillates with the core oscillations.
The anisotropy increases as the core contracts and decreases as the core expands.
One may be interested in the fact that 
the anisotropy even at 30 and 40\% radii shows a sign of oscillations,
while these radii, themselves, show no clear sign of oscillations in figure 5.
If we magnify figure 5, however, we can see that the radii actually oscillate with very small amplitudes.
That is, we can hardly see the oscillations of the 30\% and 40\% radii in figure 5, simply because their amplitudes are too small.

%\begin{center}
%------------------------------\\
%Fig. 7\\
%   vel. disp. profile, TI, N=40000
%------------------------------\\
%\end{center}

%\begin{center}
%------------------------------\\
%Fig. 8\\
%   density profile, TI, N=40000
%------------------------------\\
%\end{center}

Next, we consider a gravothermal expansion phase.
The mechanism of gravothermal oscillations was already clearly explained 
in a seminal work by Bettwieser and Sugimoto (1984).
The key feature of the gravothermal oscillations is the appearance of a temperature inversion (i.e., an outward increase of the temperature) which causes a gravothermal expansion.
Very recently, it has been clearly shown that a temperature inversion actually appears in a real $N$-body system of $N=32000$ (Makino 1996).
Figures 7 and 8 show the evolution of the velocity dispersion (or temperature) and density profiles
when a gravothermal expansion occurs in the case of $N=40000$.
At $t=18.74~\trhi$, a temperature inversion has just appeared, and the gravothermal expansion has started.
At $t=18.84~\trhi$ the amount of the temperature inversion is nearly maximum.
At this time, $\sigt^2$ slightly exceeds $\sigr^2$ in the region of the temperature hump.
This is because the radial velocity dispersion at the hump reflects a lower central temperature more than the tangential one.
The temperature inversion almost disappears at $t=19.04~\trhi$.
This indicates the end of the gravothermal expansion,
and a normal isothermal core appears again.

%\begin{center}
%------------------------------\\
%Fig. 9\\
%   density profile (halo), N=5000
%------------------------------\\
%\end{center}

Paper I showed that 
the density profile in the outer halo is approximated by a power law, 
$\rho \propto r^{-3.5}$ (cf. Spitzer, Shapiro 1972),
after the rapid development of anisotropy in the halo from the isotropic initial conditions.
Figure 9 shows the density profiles at three epochs after the core collapse for $N=5000$.
It seems that the halo density profile further approaches the power law $\rho \propto r^{-3.5}$ after the collapse.
As we can see in figure 2, 
the density profile evolves self-similarly well after the core collapse.
(Even when gravothermal oscillations occur, the halo expands nearly self-similarly.)
Therefore, in well-relaxed isolated clusters, the halo density profile is always approximated by a $r^{-3.5}$ power law.
Paper I also showed that 
the tangential velocity dispersion profile in the outer halo is reasonably
approximated by a power law, $\sigt^2 \propto r^{-2}$,
though this approximation is not as good as the approximation for the density (see figures 7 and 8 of Paper I).
This power law can be applied for post-collapse clusters as well.
Actually, the velocity dispersion profile in the halo changes little after the collapse.

\section{Conclusions and Discussion}\label{sec:conclusion}

\indent\indent

In the previous paper (Paper I) an improved numerical code for solving the orbit-averaged FP equation in energy--angular momentum space was developed
in order to study the evolution of star clusters which have {\it anisotropic} velocity distributions.
Numerical simulations were performed by using the code for the pre-collapse evolution of single-mass globular clusters in Paper I\@.
In this paper, following Paper I, the post-collapse evolution of single-mass clusters was considered.
The effect of three-body binaries was incorporated in the code as a heat source.
Actually, two partially different codes were developed in Paper I\@.
They differ in the scheme for solving the FP equation, itself:
one uses the finite-difference scheme and the other does the finite-element scheme.
The two codes have similar numerical accuracy for pre-collapse calculations.
For post-collapse calculations, however, the finite-element code is worse concerning energy conservation than the finite-difference code.
Although this difficulty of the finite-element code may be removed by using higher-order basis functions,
such efforts were not made in this study.
By using the finite-difference code we can perform post-collapse calculations with reasonable numerical accuracy (within a few percent error in energy conservation).

There is no significant difference in the evolution of the central density between the 1D and 2D FP models as far as we studied for $N=$ 5000, 10000, 20000, and 40000.
(However, there is a difference in the core collapse time, as described in Paper I;
the collapse time is a little longer in the 2D model.)
In particular, the qualitative features of gravothermal oscillations are common to the 1D and 2D models.
The appearance of a temperature inversion in the 2D model is similar to that in the 1D model.
However, a slight anisotropy appears in the region of the temperature hump:
the tangential velocity dispersion exceeds the radial one.
This is an opposite anisotropy to the usual one,
and a consequence of the lower central temperature.
The opposite anisotropy disappears along with the disappearance of the temperature inversion.

Clusters expand nearly self-similarly as a whole well after a core collapse.
In fact, the expansion is consistent with the self-similar expansion law,
$r \propto t^{2/3}$.
The core-halo structure is more developed in the 2D model than in the 1D model.
However, the evolution of the half-mass radius in the two models is almost identical, if the time of one model is scaled so that the collapse times in the two models should coincide.
The density profile in the outer halo is approximated by a power law, 
$\rho \propto r^{-3.5}$, after the core collapse as well as before it.

The anisotropy at the inner Lagrangian radii decreases during a rapid core-expansion phase just after the core bounce.
When the core expansion is stable (e.g. for $N=5000$), the anisotropy at each of the inner radii settles to a roughly constant value, because the inner radii expand self-similarly as well as the half-mass and outer radii.
When  gravothermal oscillations occur (e.g. for $N=$ 20000, 40000), the anisotropy at the inner radii oscillates with the core oscillations.
The anisotropy at the outer Lagrangian radii continues to increase slowly after the core bounce, whether the core oscillations occur or not.
The rate of anisotropy increase at the outer radii slows down as the cluster expands.
This is mainly because the central relaxation time gets longer.
If we measure time in units of the central relaxation time [see equation (\ref{eq:tau})], 
the anisotropy increase rate at the outer radii is almost constant, except for the initial epochs of the calculations when the anisotropy increases very rapidly.

There are other currently-working numerical codes which can deal with the
velocity anisotropy:
one of them is Spurzem's code, which is based on the anisotropic gaseous model (e.g. Spurzem 1996);
the other is Giersz's code, which solves the FP equation by a Monte-Carlo technique (Giersz 1996).
On the other hand, Giersz and Heggie (1994a, b) showed that 
the combination of a large number of $N$-body simulations leads to results of high statistical quality 
which can give valuable information concerning the theory of stellar dynamics.
Some comparisons between $N$-body, isotropic/anisotropic gaseous, and isotropic FP models were made for isolated one- or two-component clusters
(Giersz, Heggie 1994a, b; Giersz, Spurzem 1994; Spurzem, Takahashi 1995).
Those comparisons showed that the results of the FP and gaseous models are generally in good agreement with the statistical data of $N$-body simulations.
However, differences between the statistical models and the $N$-body models remain in some other respects.
2D FP models may give a better agreement with $N$-body models.
Comparisons of the 2D FP models with the anisotropic gaseous and $N$-body models are now in progress.
A preliminary result of such comparisons for $N=10000$ models is presented by Spurzem (1996) and Takahashi (1996).
For example, concerning the evolution of anisotropy in the halo,
the agreement between the 2D FP and $N$-body models is very good.
This fact supports the reliability of our 2D FP models.

Through Paper I and this paper, we have investigated the evolution of {\it realistic} anisotropic models of globular clusters.
However, these models are {\it unrealistic} in some respects.
They do not consider the stellar-mass spectrum, the galactic tidal field, and the effects of tidal and primordial binaries.
The stellar mass-loss may also have an important effect on the initial evolutionary stages of clusters (Chernoff, Weinberg 1990).
More realistic models incorporating some or all of these various effects will be studied in the future.

\bigskip\bigskip
%Acknowledgements

I would like to thank Professor S. Inagaki for valuable comments.
I also wish to thank the referee, Professor H. Cohn, for useful comments which greatly helped to improve the presentation of the paper.
This work was supported in part by the Grand-in-Aid for Encouragement of Young
Scientists by the Ministry of Education, Science, Sports and Culture of Japan 
(No. 1338).

%\clearpage

\begin{reference}
\item Bettwieser E., Sugimoto D. 1984, \mn\ 208, 493
\item Breeden J.L., Cohn H.N., Hut P., 1994, \apj\ 421, 195
\item Breeden J.L., Cohn H.N., 1995, \apj\ 448, 672
\item Chang J.S., Cooper G. 1970, J. Comp. Phys. 6, 1
\item Chernoff D.F., Weinberg M.D. 1990, \apj\ 351, 121
\item Cohn H. 1979, \apj\ 234, 1036
\item Cohn H. 1980, \apj\ 242, 765
\item Cohn H. 1985, in Dynamics of Star Clusters, IAU Symp No.113, ed
J.~Goodman, P.~Hut (D.~Reidel Publishing Company, Dordrecht) p161
\item Cohn H., Hut P., Wise M., 1989, \apj\ 342, 814
\item Drukier G.A. 1995, \apjs\ 100, 347
\item Giersz M. 1996, in Dynamical Evolution of Star Clusters, IAU Symp No.174, ed P.~Hut, J.~Makino (Kluwer, Dordrecht) in press
\item Giersz M., Heggie D.C. 1994a, \mn\ 268, 257
\item Giersz M., Heggie D.C. 1994b, \mn\ 270, 298
\item Giersz M., Spurzem R. 1994, \mn\ 269, 241
\item Goodman J. 1984, \apj\ 280, 298
\item Goodman J. 1987, \apj\ 313, 576
\item Heggie D.C., Ramamani N. 1989, \mn\ 237, 757
\item H\'enon M. 1965, Ann. Astrophys. 28, 62
\item Hut P. 1985, in Dynamics of Star Clusters, IAU Symp No.113, ed
J.~Goodman, P.~Hut (D.~Reidel Publishing Company, Dordrecht) p231
\item Inagaki S., Lynden-Bell D. 1990, \mn\ 244, 254
\item Louis P.D. 1990, \mn\ 244, 478
\item Makino J. 1996, in Dynamical Evolution of Star Clusters, IAU Symp No.174, ed P.~Hut, J.~Makino (Kluwer, Dordrecht) in press
\item Spitzer L.Jr, 1987, Dynamical Evolution of Globular Clusters (Princeton
University Press, Princeton)
\item Spitzer L.Jr, Shapiro S.L. 1972, \apj\ 173, 529
\item Spurzem R. 1994, in Ergodic Concepts in Stellar Dynamics, 
ed  D. Pfenniger, V.A. Gurzadyan (Springer, Berlin) p170
\item Spurzem R. 1996, in Dynamical Evolution of Star Clusters, IAU Symp No.174, ed P.~Hut, J.~Makino (Kluwer, Dordrecht) in press
\item Spurzem R., Takahashi K. 1995, \mn\ 272, 772
\item Takahashi K. 1993, \pasj\ 45, 789
\item Takahashi K. 1995, \pasj\ 47, 561 (Paper I)
\item Takahashi K. 1996, in Dynamical Evolution of Star Clusters, IAU Symp No.174, ed P.~Hut, J.~Makino (Kluwer, Dordrecht) in press

\end{reference}

\clearpage

\begin{figure}[htb]
\begin{center}
\leavevmode
\epsfverbosetrue
\epsfxsize=15cm \epsfbox{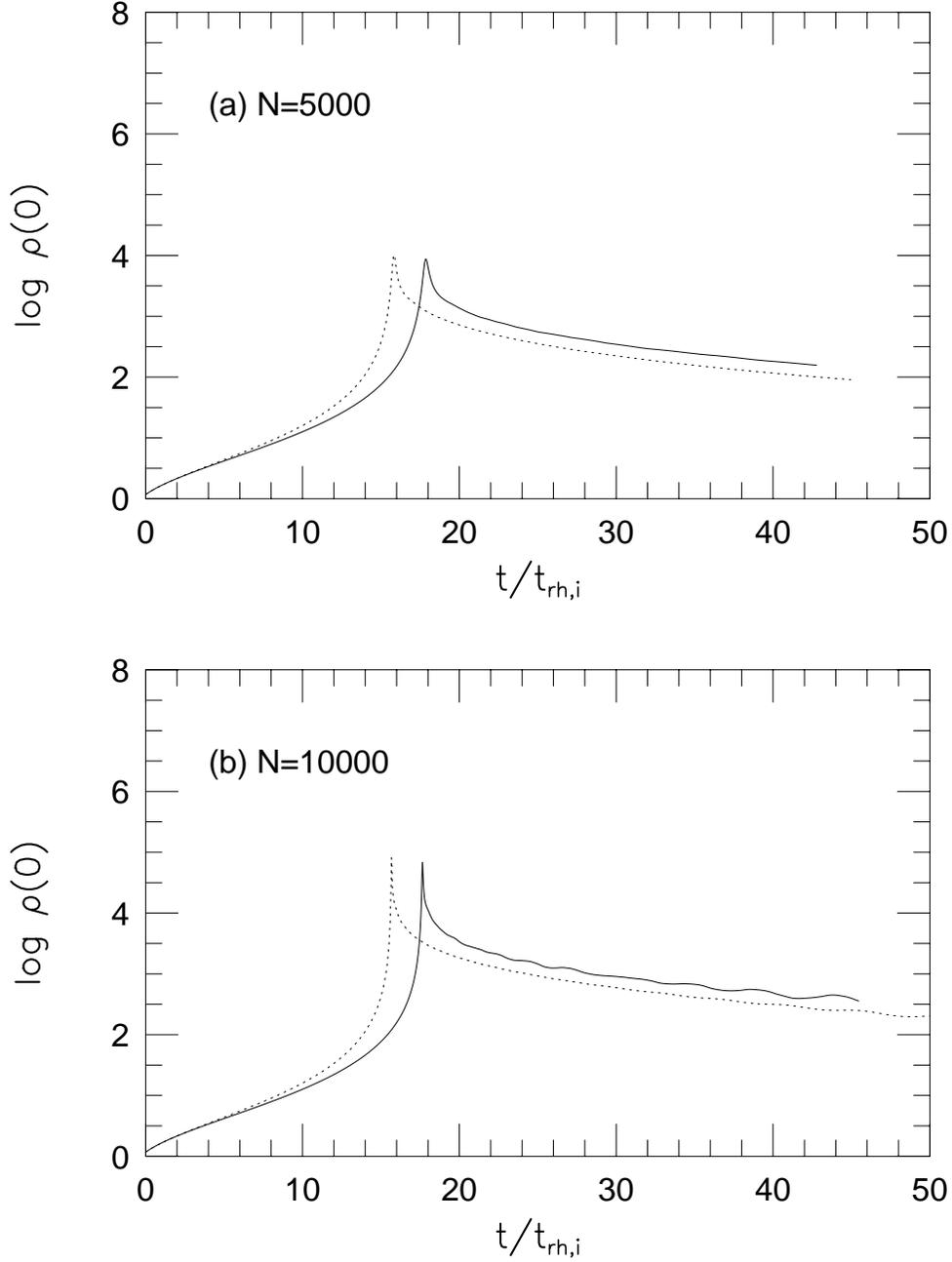}
\end{center}
\caption{
Evolution of the central density for (a) $N=5000$, (b) $N=10000$, (c)
(c) $N=20000$, and (d) $N=40000$.
The solid curves are the results of the 2D FP calculations, and the
dotted curves are the results of the 1D FP calculations.
The time is measured in units of the initial half-mass relaxation time
$t_{\rm rh,i}$.
}
\end{figure}

\clearpage

\setcounter{figure}{0}
\begin{figure}[htb]
\begin{center}
\leavevmode
\epsfverbosetrue
\epsfxsize=15cm \epsfbox{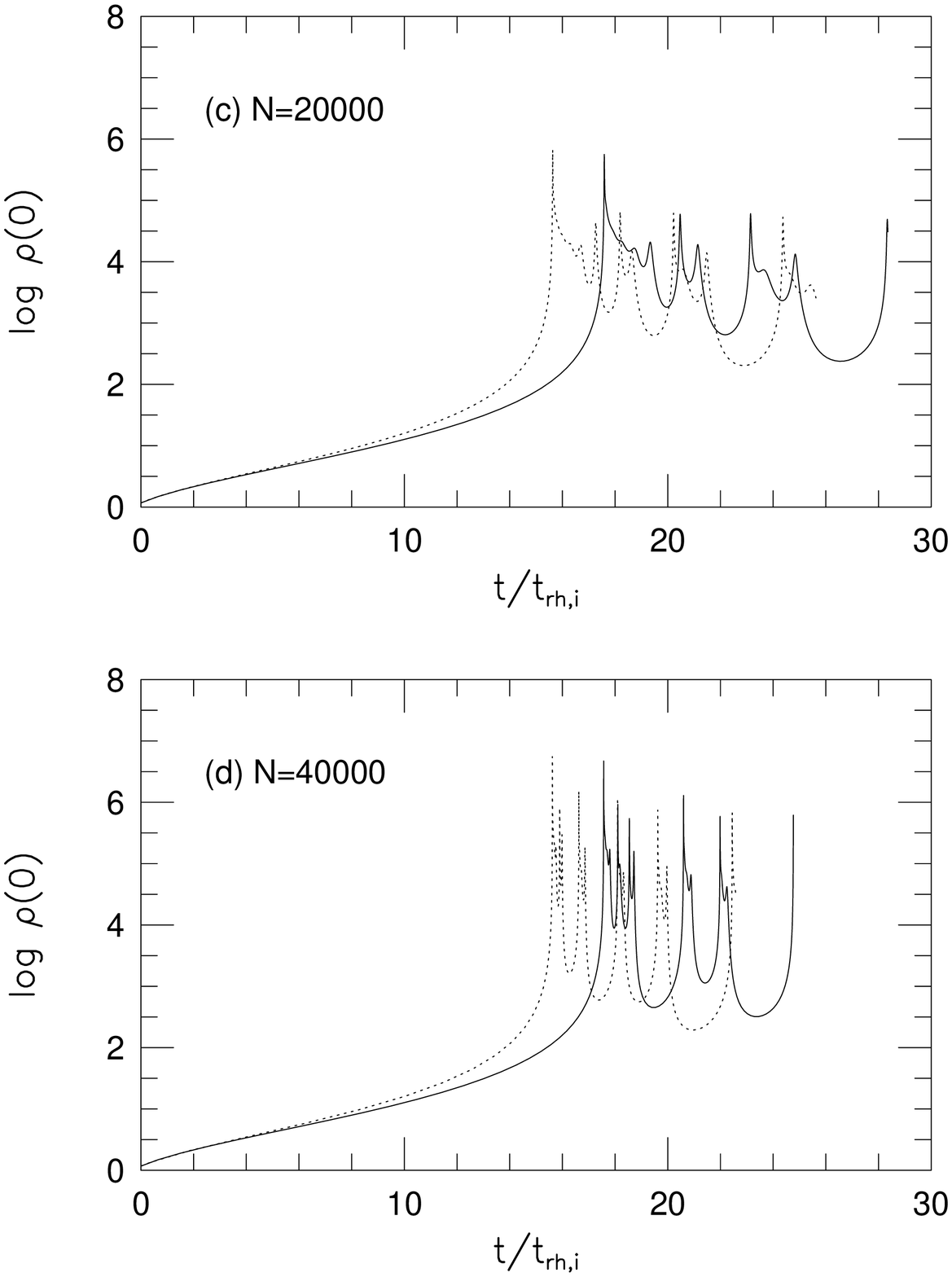}
\end{center}
\caption{
{\it continued}
}
\end{figure}

\clearpage

\begin{figure}[htb]
\begin{center}
\leavevmode
\epsfverbosetrue
\epsfxsize=15cm \epsfbox{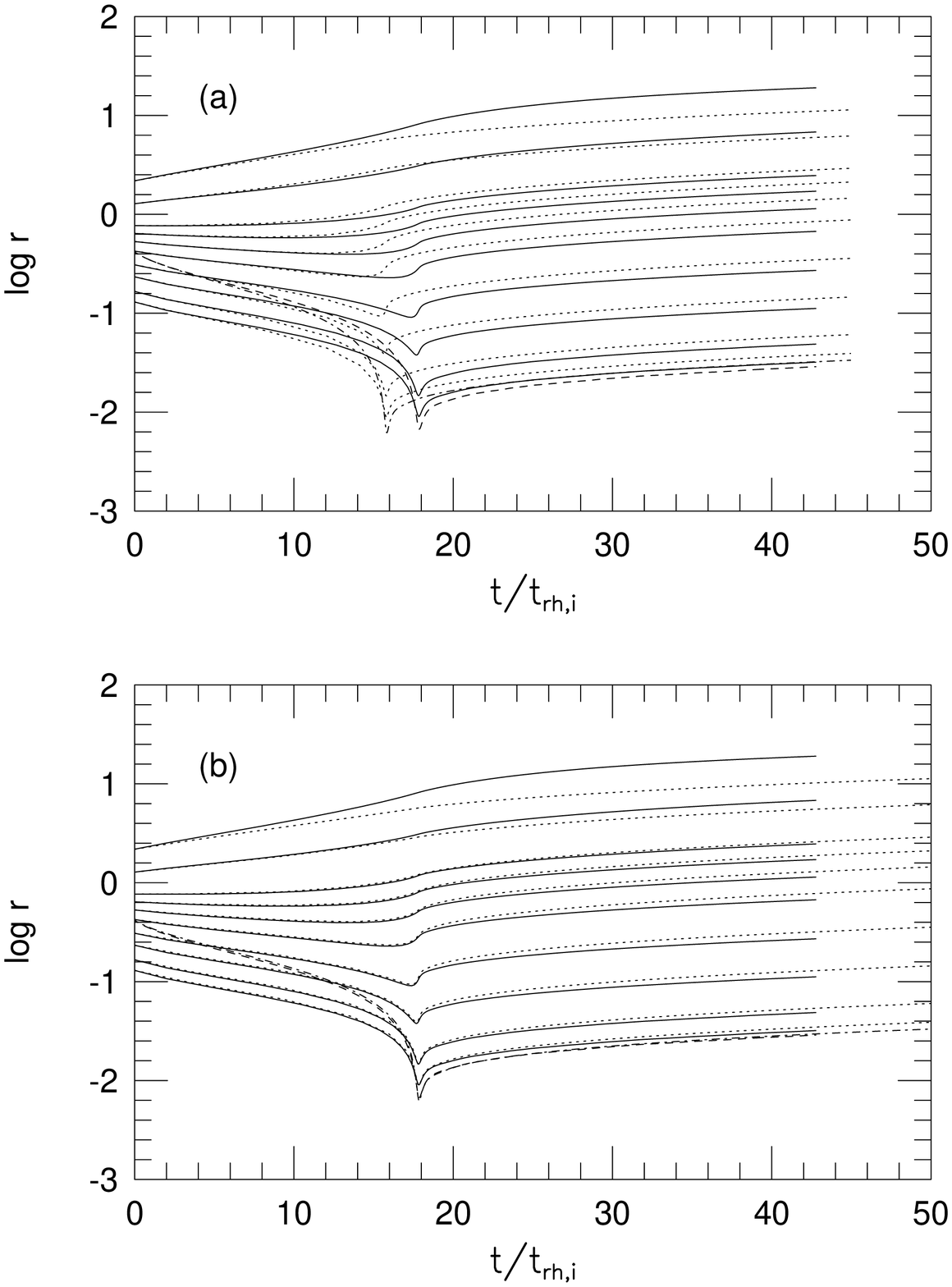}
\end{center}
\caption{
(a) Evolution of Lagrangian radii containing
1, 2, 5, 10, 20, 30, 40, 50, 75, and 90\% of the total mass, for
$N=5000$.
The solid curves are the result of the 2D FP calculation, and the dotted
curves are the result of the 1D FP calculation.
The core radii are also plotted by the dashed curve (2D) and the
dash-dotted curve (1D).
(b) Same as (a), but
the time of the 1D calculation is scaled
so that the collapse time in the 1D calculation should coincide with
that in the 2D calculation.
}
\end{figure}

\clearpage

\begin{figure}[htb]
\begin{center}
\leavevmode
\epsfverbosetrue
\epsfxsize=15cm \epsfbox{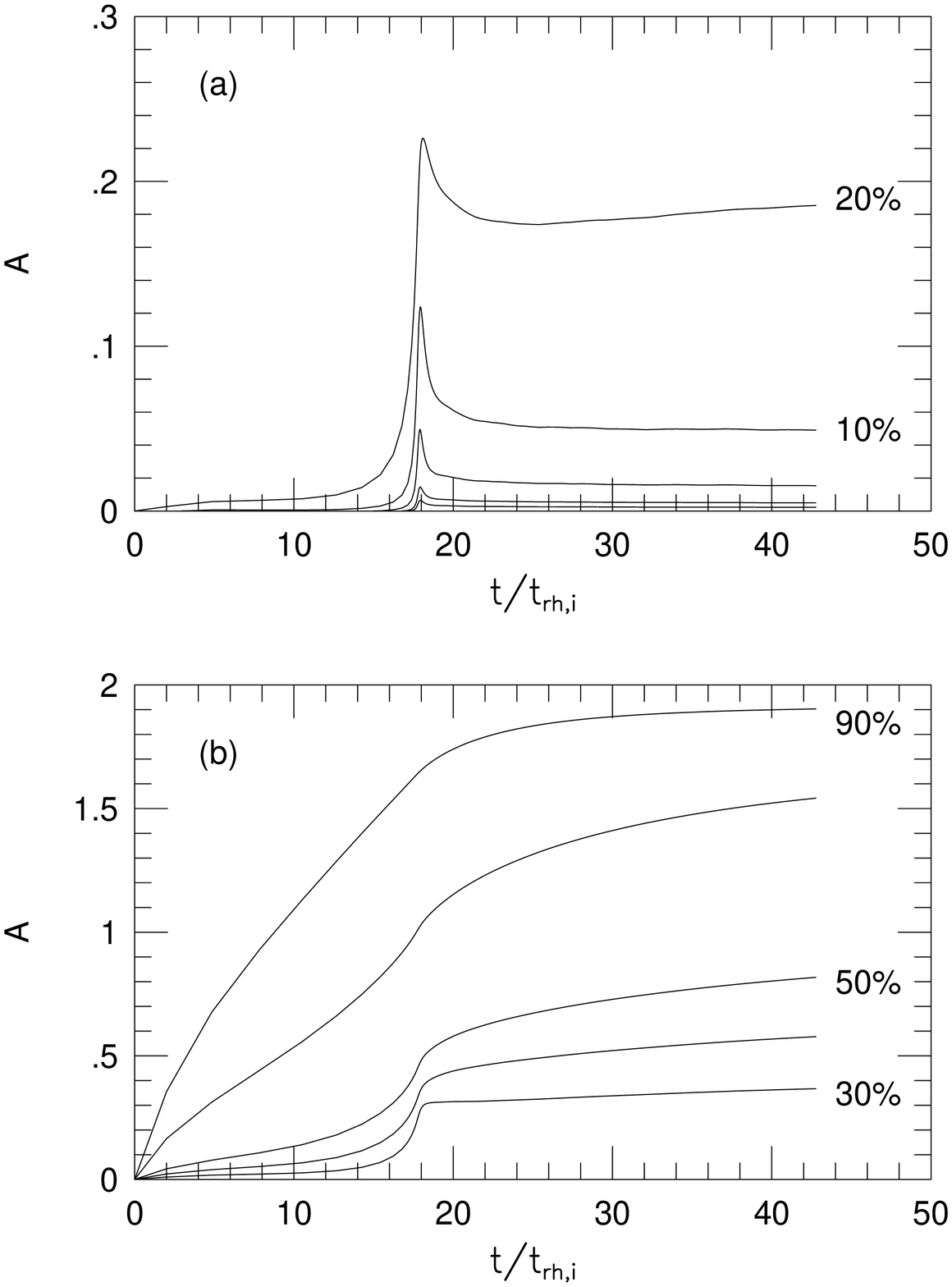}
\end{center}
\caption{
Evolution of the anisotropy parameter,
$A \equiv 2-2\sigma_{\rm t}^2/\sigma_{\rm r}^2$,
at the (a) inner (1--20\%) and (b) outer (30--90\%) Lagrangian radii,
for $N=5000$.
}
\end{figure}

\clearpage

\begin{figure}[htb]
\begin{center}
\leavevmode
\epsfverbosetrue
\epsfxsize=15cm \epsfbox{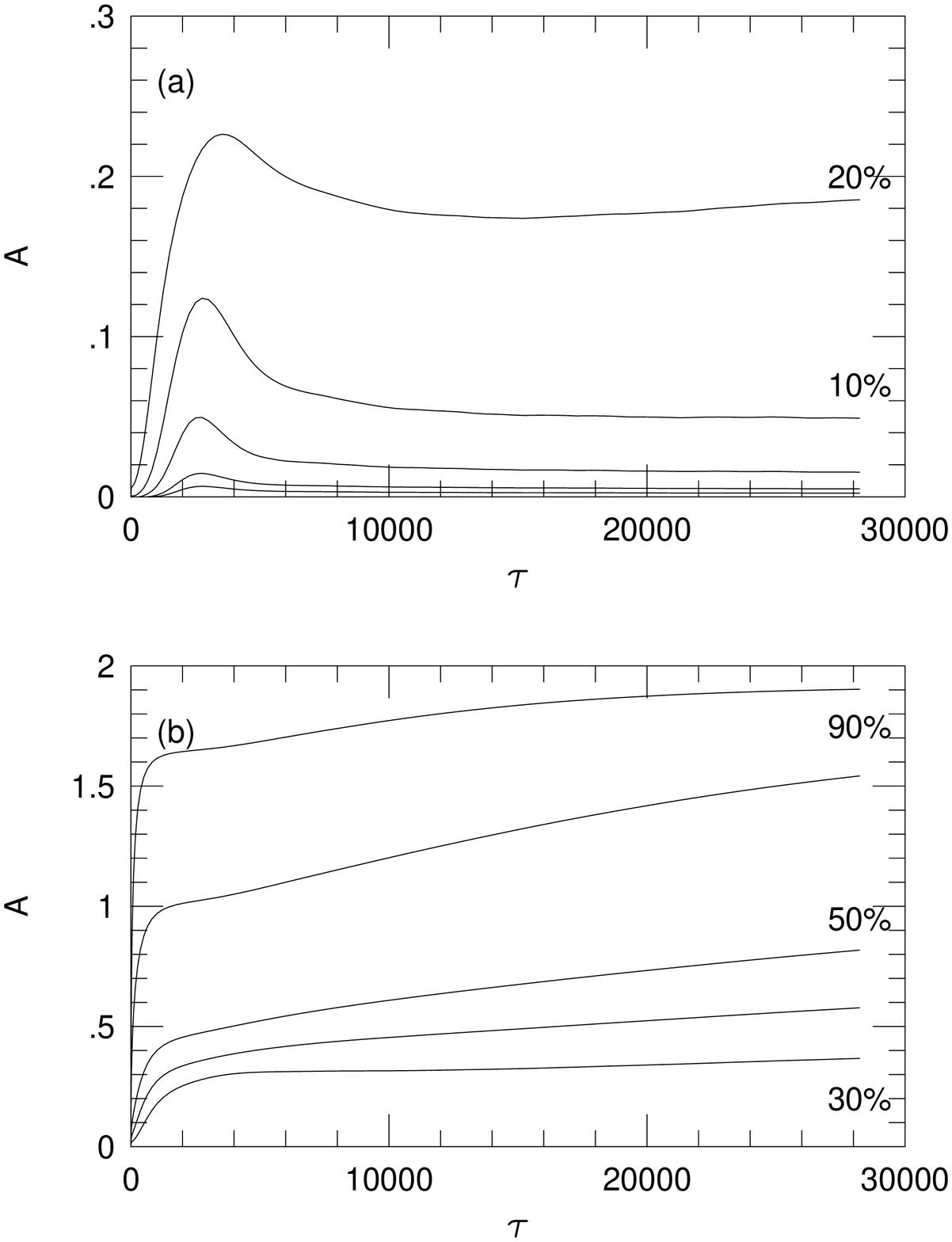}
\end{center}
\caption{
Same as figure 3, but the abscissa is the elapsed number of actual
central relaxation times, $\tau$ [see equation (5)].
The core bounce occurs at $\tau \approx 2400$.
}
\end{figure}

\clearpage

\begin{figure}[htb]
\begin{center}
\leavevmode
\epsfverbosetrue
\epsfxsize=15cm \epsfbox{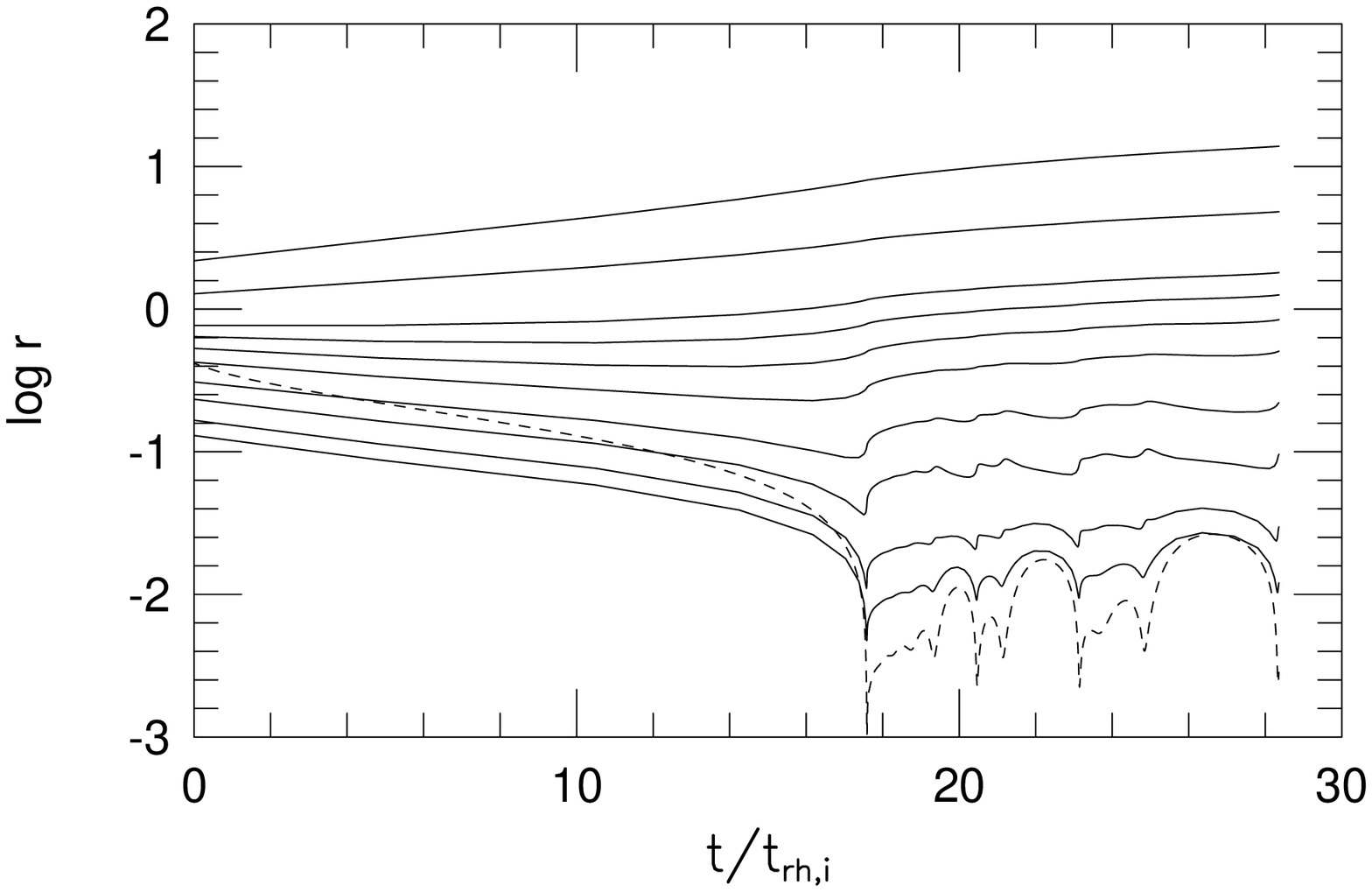}
\end{center}
\caption{
Evolution of the Lagrangian radii in the 2D model for $N=20000$.
The dashed curve represents the core radius.
}
\end{figure}

\clearpage

\begin{figure}[htb]
\begin{center}
\leavevmode
\epsfverbosetrue
\epsfxsize=15cm \epsfbox{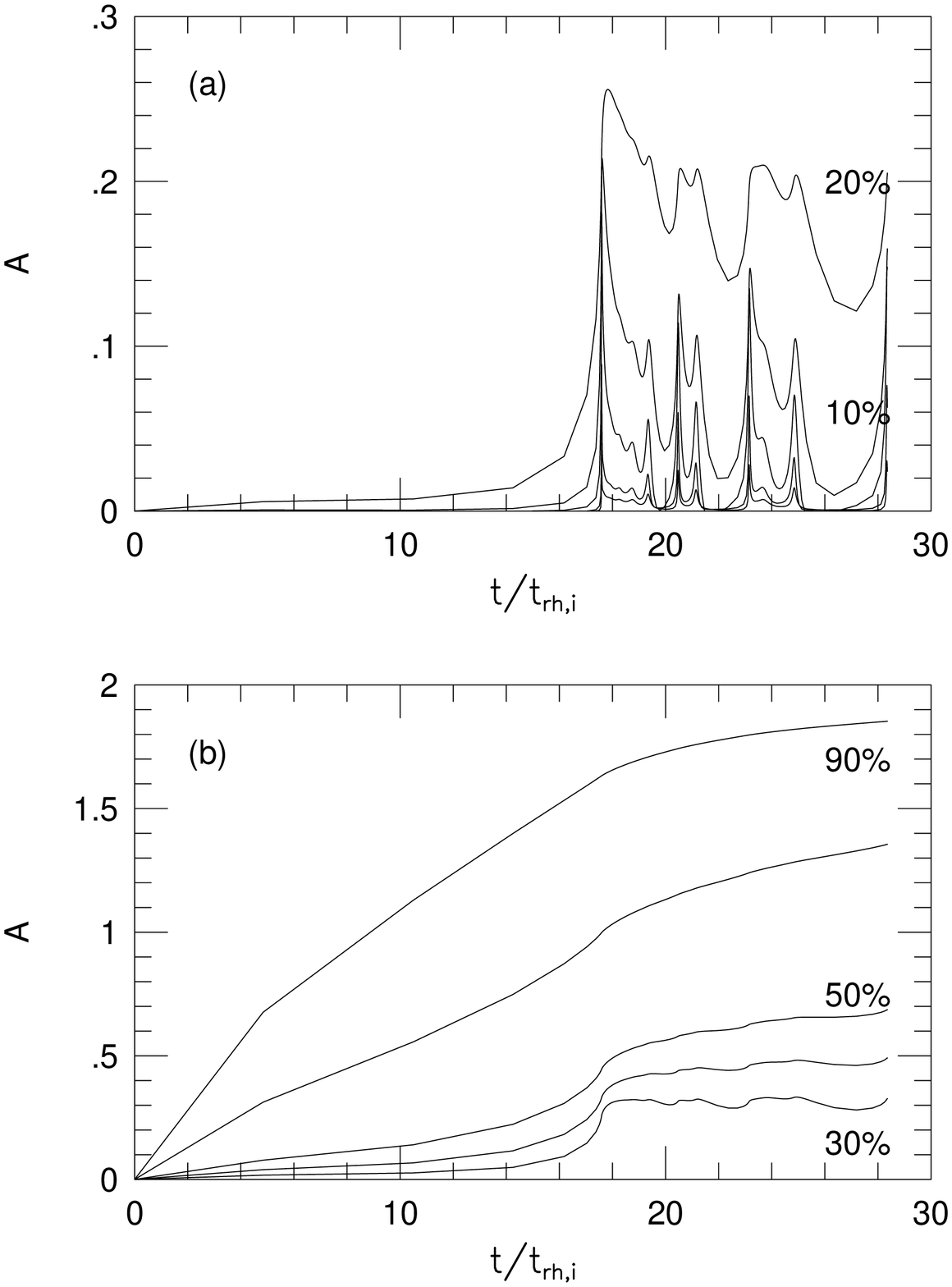}
\end{center}
\caption{
Same as figure 3, but for $N=20000$.
}
\end{figure}

\clearpage

\begin{figure}[htb]
\begin{center}
\leavevmode
\epsfverbosetrue
\epsfxsize=15cm \epsfbox{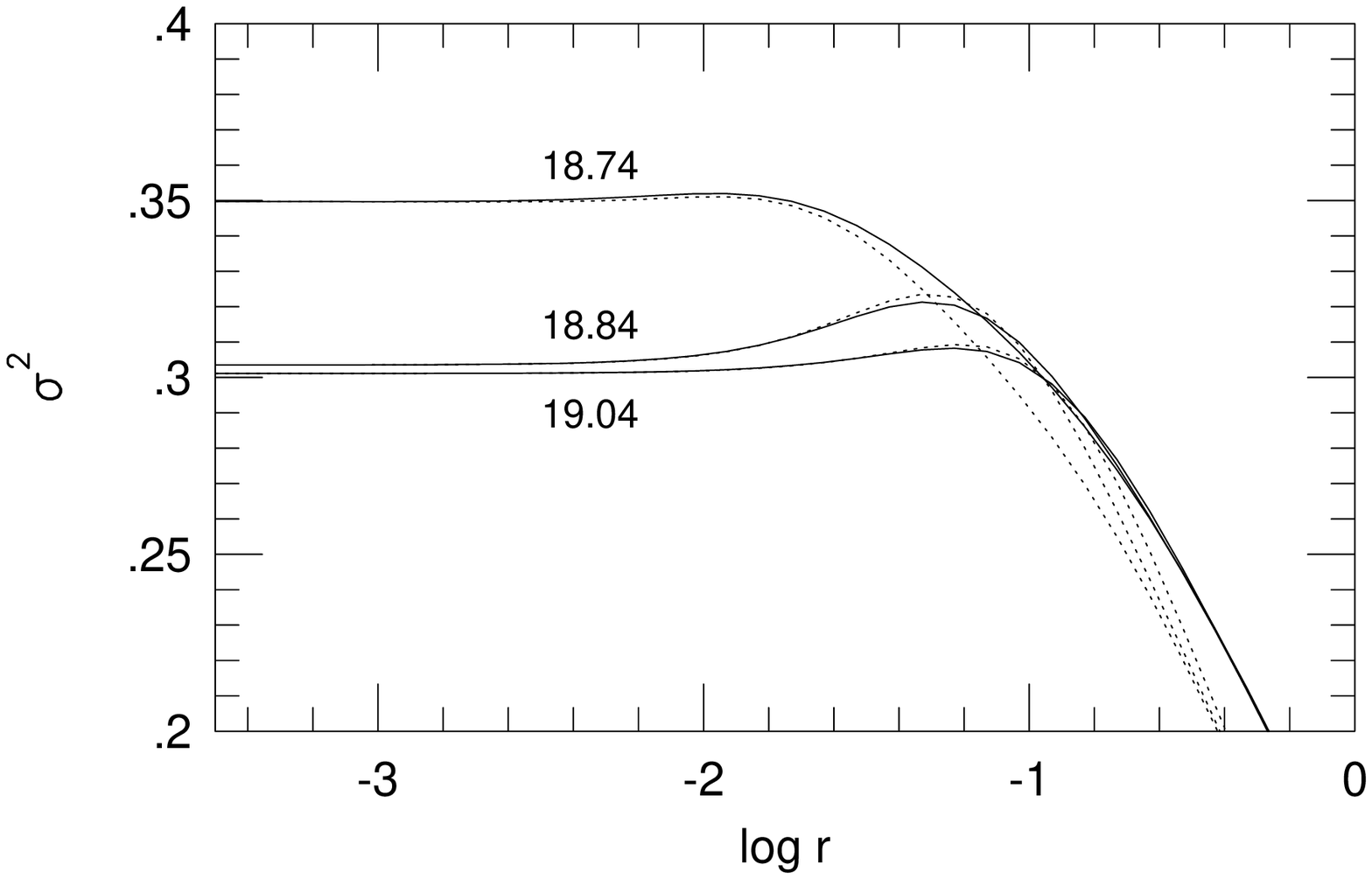}
\end{center}
\caption{
Evolution of the velocity dispersion (or temperature) profile when a
temperature inversion appears (the times are indicated in the figure
in units of the initial half-mass relaxation time),
in the 2D FP model for $N=40000$.
The solid and dotted curves are the radial and tangential velocity
dispersions, respectively.
}
\end{figure}

\clearpage

\begin{figure}[htb]
\begin{center}
\leavevmode
\epsfverbosetrue
\epsfxsize=15cm \epsfbox{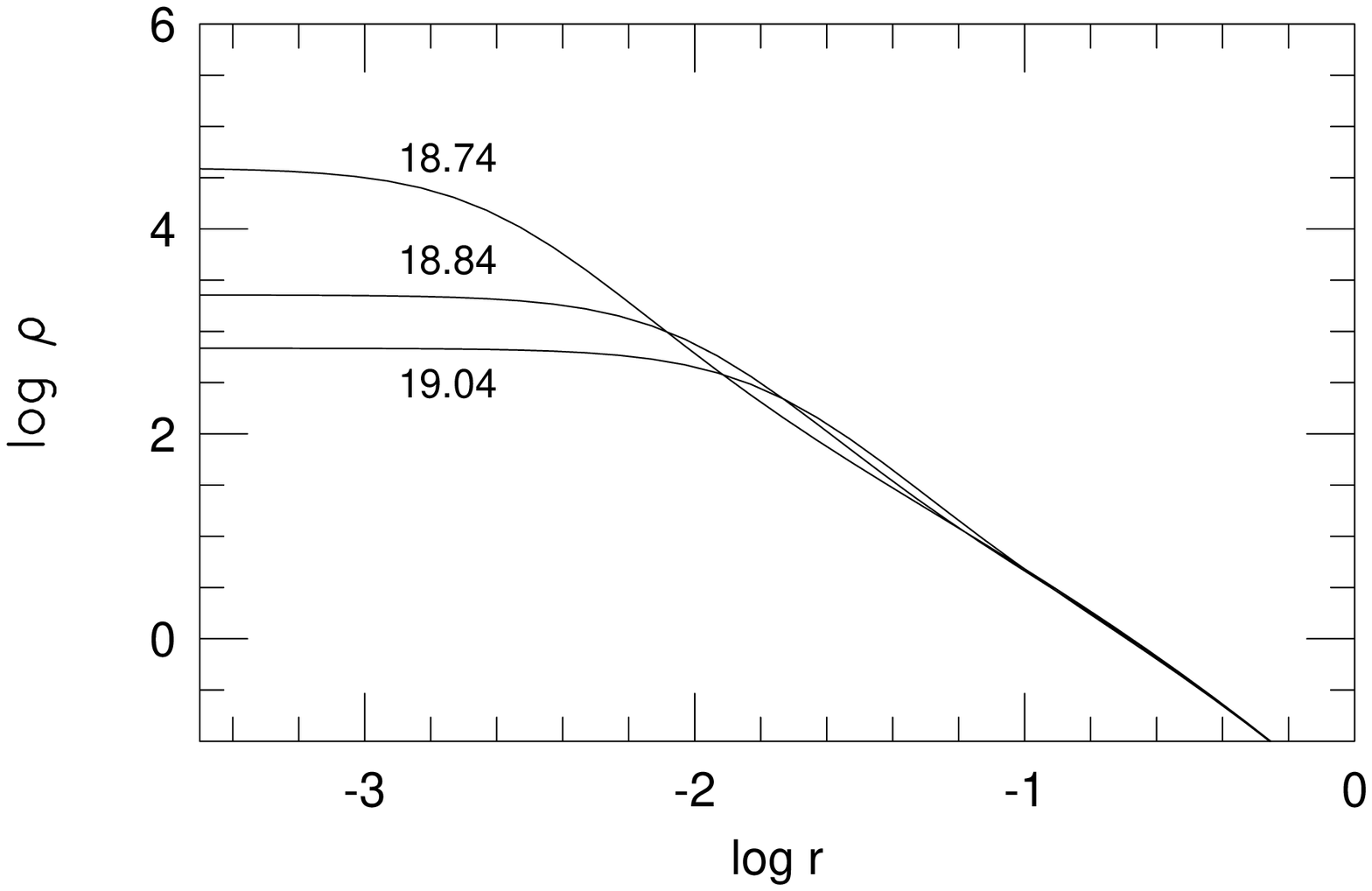}
\end{center}
\caption{
Evolution of the density profile which corresponds to the velocity
dispersion profile shown in figure 7.
}
\end{figure}

\clearpage

\begin{figure}[htb]
\begin{center}
\leavevmode
\epsfverbosetrue
\epsfxsize=15cm \epsfbox{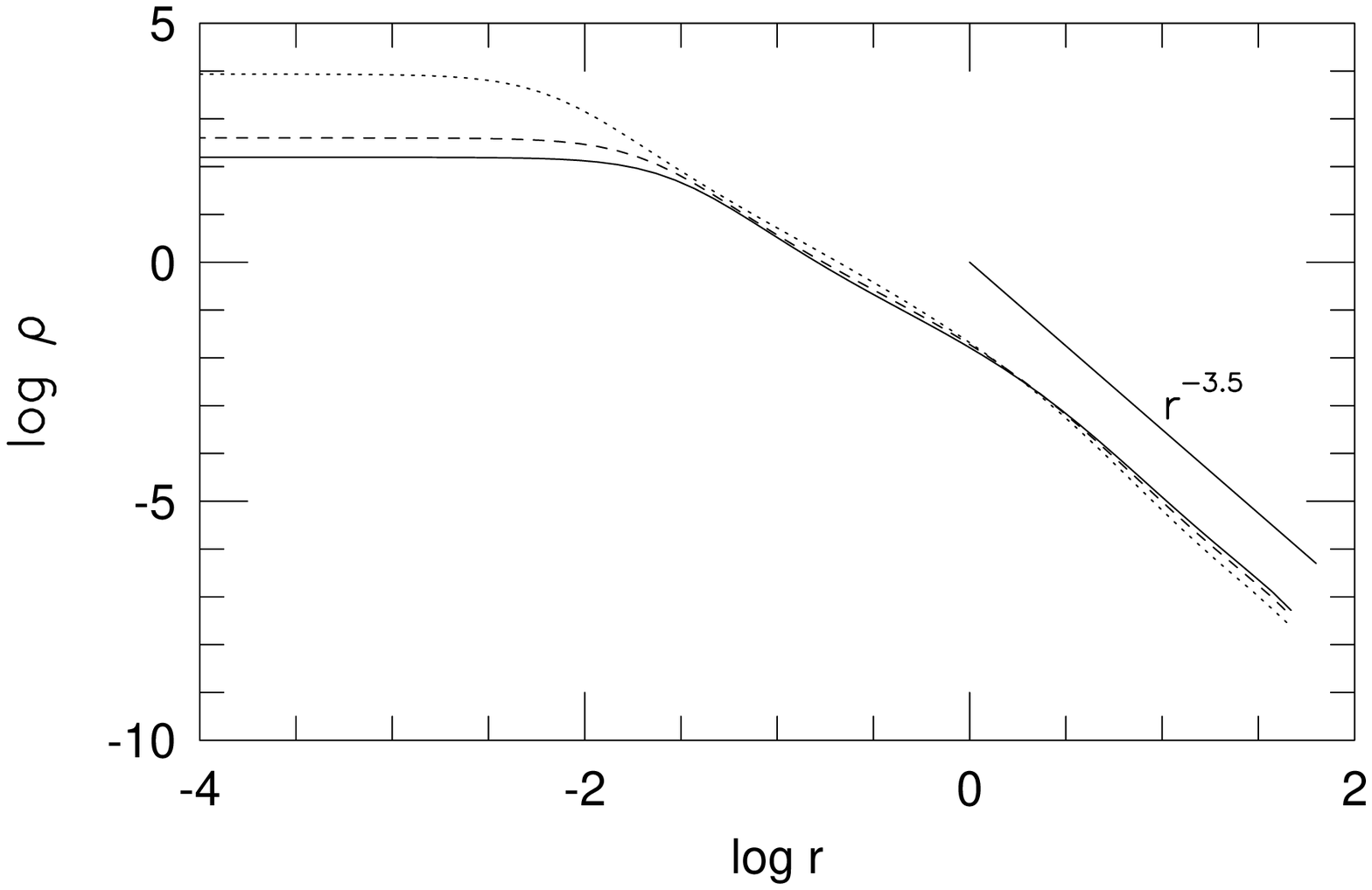}
\end{center}
\caption{
Density profiles at three epochs after the core collapse in the 2D FP
model for $N=5000$.
The dotted, dashed, and solid curves represent the profiles
at $t/\trhi=$ 17.9 (just after the collapse time),  28.4, and 42.8,
respectively.
The asymptotic line $\rho \propto r^{-3.5}$ is shown for a comparison.
}
\end{figure}

\end{document}